\def\comment#1{}

\def\cm#1{}

\documentstyle[12pt,pra,epsf,aps]{revtex}

\begin{document}
\title{Smearing Formula for
Higher-Order Effective Classical Potentials}
\author{Hagen Kleinert, Werner K\"urzinger and Axel Pelster\\
        Institut f\"ur Theoretische Physik, Freie Universit\"at Berlin,\\
        Arnimallee 14, D--14195 Berlin, Germany\\ $\mbox{}$\\}
\maketitle
\begin{abstract}
In the variational approach to quantum statistics,
a smearing
formula describes efficiently the consequences of quantum fluctuations
upon an interaction potential.
The result is an effective classical
potential from which the partition function
can be obtained by a simple integral.
In this work, the smearing formula
is extended to higher orders in the variational perturbation theory.
An application to
the singular Coulomb potential
exhibits
the same
fast convergence
with
increasing orders that
has been observed
in previous variational perturbation expansions
of the anharmonic oscillator with quartic potential.
\end{abstract}
\section{Introduction}
The variational approach to quantum statistics,
initially
based on the Jensen-Peierls inequality for
imaginary-time path integrals
\cite{Feynman1,Feynman2}, yielded
crude upper bounds
for the free
energy of many quantum mechanical systems at
all temperatures and coupling strengths.
By abandoning the inequality,
the approach has been extended \cite{Kleinert3} to a
systematic, fast convergent variational perturbation theory
\cite{Kleinert4,Meyer,Kleinert2},
within which the original approach is just a first-order
approximation.
Thermodynamic and some local  quantities
can now be evaluated
to any desired accuracy, starting out from an ordinary perturbation
expansion of arbitrary order.\\[-3mm]

A particularly attractive feature of the original
variational approach
was the existence of a
smearing formula
in the form of a
Gaussian convolution integral
which compactly accounts for the effect of quantum fluctuations
upon the interaction potential
and other local quantities \cite{Feynman2,Kleinert2}.
This
formula was applicable to some classes
of singular potentials such as the Coulomb potential \cite{Kleinert2,Janke}.
There is a definite need for
such a formula
in higher orders of variational perturbation theory,
which
so far was based
on Feynman diagrams, thus being
limited to
polynomial interactions.
The purpose of this paper is to derive
the desired higher-order
smearing formula. This will be done
in Section III after a
brief review
of variational perturbation theory in
Section II.
An application to the Coulomb potential is
given
in Section IV, where the
effective classical potential is calculated to
second order
in the variational perturbation theory.
Its zero-temperature limit yields in Section V a
variational perturbation expansion for the ground
state energy up to second order. Section VI reproduces
this result by a direct variational treatment of the
Rayleigh-Schr\"odinger perturbation expansion, and carries it
to third order to demonstrate
the fast convergence of the variational perturbation expansion.
\section{Review of Variational Perturbation Theory}
Consider
a quantum mechanical
point particle of mass $M$ moving
in a one-dimensional time-independent
potential $V ( x )$. Its
thermodynamic partition function is given by the
imaginary-time path integral \cite{Kleinert2}
\begin{equation}
\label{PF1}
Z = \oint {\cal D} x ( \tau ) \exp \left\{ - \frac{1}{\hbar}
{\cal A} [ x ( \tau ) ] \right\} \, ,
\end{equation}
with the euclidean action
\begin{equation}
\label{A}
{\cal A} [ x ( \tau ) ] = \int\limits_0^{\hbar \beta}
d \tau \left[ \frac{M}{2} \dot{x} ( \tau )^2 + V ( x ( \tau ) ) \right] \, ,
\end{equation}
and the abbreviation
$\beta \equiv 1/k_B T$. The
paths $x ( \tau )$
satisfy the periodic boundary condition
$x ( 0 ) = x ( \hbar \beta )$.
Following Feynman \cite{Feynman1}, we decompose
the path integral for the partition function (\ref{PF1}) into an
ordinary integral over the time-averaged position
\begin{equation}
\label{XE}
x_0=\overline{x} \equiv
\frac{1}{\hbar\beta} \int\limits_0^{\hbar\beta} d \tau \, x ( \tau ) \, ,
\end{equation}
and a remaining path integral over the
fluctuations
\begin{equation}
 \delta x(\tau )=x(\tau )-x_0
\label{devia}\end{equation}
 around
$x_0$.
Thus we
rewrite (\ref{PF1}) as an integral
\begin{equation}
\label{PF2}
Z =  \int\limits_{- \infty}^{+ \infty}
\tilde d x_0 \, Z^{x_0} \,
\end{equation}
over a {\em local partition function\/} $Z^{x_0}$
 which is
defined by the restricted path integral
\begin{equation}
\label{LPF}
Z^{x_0} \equiv \oint {\cal D} x ( \tau ) \, \tilde\delta
( \overline{x} - x_0 ) \exp
\left\{ - \frac{1}{\hbar} {\cal A} [ x ( \tau ) ] \right\} \, ,
\end{equation}
with the notation
\begin{equation}
\tilde dx_0\equiv \sqrt{\frac{M}{2 \pi \hbar^2 \beta}} \, dx_0\, , \hspace*{2cm}
\tilde  \delta(\bar x-x_0)\equiv  \sqrt{\frac{2 \pi \hbar^2 \beta}{M}}
\, \delta ( \overline{x} - x_0 ).
\label{@S}\end{equation}
The free energy associated with the local partition function
(\ref{LPF}) is defined as the
{\em effective classical potential\/} \cite{Kleinert2}
\begin{equation}
\label{ECP}
V^{{\rm eff, cl}} ( x_0 ) = - \frac{1}{\beta} \, \log  Z^{x_0}
\end{equation}
which accounts for the effects of all quantum fluctuations.\\[-3mm]

In order to calculate
$V^{{\rm eff, cl}} ( x_0 )$, we decompose the euclidean action (\ref{A}) into a sum
\begin{equation}
\label{DECO}
{\cal A} [ x ( \tau ) ] = {\cal A}^{x_0}_{\Omega} [ x ( \tau ) ]
+ {\cal A}^{x_0}_{{\rm int}} [ x ( \tau ) ] \, ,
\end{equation}
where the first term
is the action of a harmonic oscillator centered around $x_0$
with an undetermined local {\em trial frequency\/} $\Omega ( x_0 )$,
\begin{equation}
\label{TA}
{\cal A}^{x_0}_{\Omega} [ x ( \tau ) ] = \int\limits_0^{\hbar \beta}
 d \tau \left\{ \frac{M}{2} \dot{x} ( \tau )^2 + \frac{M}{2} \Omega^2
( x_0 ) \left[ x ( \tau ) - x_0 \right]^2 \right\} ,
\end{equation}
and the second term is the remaining interaction
\begin{equation}
{\cal A}^{x_0}_{{\rm int}} [ x ( \tau ) ] = \int\limits_0^{\hbar\beta}
d \tau \, V^{x_0}_{{\rm int}} ( x ( \tau ) )
\end{equation}
of the potential difference
\begin{equation}
V^{x_0}_{{\rm int}} ( x ) = V ( x ) - \frac{M}{2} \Omega^2 ( x_0 )
( x - x_0 )^2 \, .
\label{@intpot}
\end{equation}
With this decomposition,
the local partition function
(\ref{LPF}) may be expanded in powers of
the interaction potential (\ref{@intpot})
around
the local harmonic partition function
\begin{equation}
\label{LHPF}
Z^{x_0}_{\Omega} =
 \oint {\cal D} x ( \tau ) \, \tilde\delta ( \overline{x} - x_0 ) \exp
\left\{ - \frac{1}{\hbar} {\cal A}^{x_0}_{\Omega} [ x ( \tau ) ] \right\} \, .
\end{equation}
The expansion
reads
\begin{eqnarray}
\hspace*{0.5cm}
Z^{x_0} & = & Z^{x_0}_{\Omega} \left\{ 1 - \frac{1}{\hbar}
\int\limits_0^{\hbar
\beta} d \tau_1 \langle V^{x_0}_{{\rm int}} ( x ( \tau_1 ) )
\rangle^{x_0}_{\Omega}
+  \frac{1}{2 \hbar^2}
\int\limits_0^{\hbar\beta} d \tau_1 \int\limits_0^{\hbar\beta} d \tau_2
\langle V^{x_0}_{{\rm int}} ( x ( \tau_1 ) )
V^{x_0}_{{\rm int}} ( x ( \tau_2 ) ) \rangle^{x_0}_{\Omega}
\right. \nonumber \\
&& ~~~~~~~~~-  \left. \frac{1}{6 \hbar^3}
\int\limits_0^{\hbar\beta} d \tau_1 \int\limits_0^{\hbar\beta} d \tau_2
\int\limits_0^{\hbar\beta} d \tau_3
\langle V^{x_0}_{{\rm int}} ( x ( \tau_1 ) )
V^{x_0}_{{\rm int}} ( x ( \tau_2 ) ) V^{x_0}_{{\rm int}} ( x ( \tau_3 ) )
\rangle^{x_0}_{\Omega}
+ \ldots \right\} \, ,
\label{PE}
\end{eqnarray}
where
the $x_0$-dependent expectation
values $\langle F_1 ( x ( \tau_1 ) ) \ldots F_n ( x ( \tau_n ) )
\rangle^{x_0}_{\Omega}$
are
correlation functions of the
local
harmonic trial system:
\begin{eqnarray}
&&\!\!\!\!\!\!\!\!\!\!\langle F_1 ( x ( \tau_1 ) ) \cdots F_n ( x ( \tau_n ) )
\rangle^{x_0}_{\Omega} = \frac{1}{Z^{x_0}_{\Omega}}
  \hspace*{2cm} \nonumber \\
&&~~~~~\times \, \oint {\cal D} x ( \tau )
F_1 ( x ( \tau_1 ) ) \cdots F_n ( x ( \tau_n ) )
\tilde\delta ( \overline{x} - x_0 ) \exp
\left\{ - \frac{1}{\hbar} {\cal A}^{x_0}_{\Omega} [ x ( \tau ) ] \right\}
\, .\label{HEV}
\end{eqnarray}
The correlation functions
can be decomposed into connected ones
via the
standard cumulant expansion
\cite{Kleinert2,Risken},
yielding for the effective classical potential
$
V^{{\rm eff,cl}} ( x_0 )
$
the following perturbation expansion
\cite{Kleinert2}
\begin{eqnarray}
~~\!\!V^{{\rm eff,cl}} ( x_0 ) & = &
F^{x_0}_ \Omega
+ \frac{1}{\hbar\beta}
\int\limits_0^{\hbar\beta} \!d \tau_1  \!
\, \langle V^{x_0}_{{\rm int}} ( x ( \tau_1 ) )
\rangle^{x_0}_{\Omega}
\!-\!\frac{1}{2 \hbar^2\beta}
\int\limits_0^{\hbar\beta}\! d \tau_1 \!\int\limits_0^{\hbar\beta}\! d \tau_2
\langle V^{x_0}_{{\rm int}} ( x ( \tau_1 ) ) 
V^{x_0}_{{\rm int}} ( x ( \tau_2 ) ) \rangle^{x_0}_{\Omega,c}
\nonumber \\
&&+\frac{1}{6\hbar^3\beta}
\int\limits_0^{\hbar\beta} d \tau_1 \int\limits_0^{\hbar\beta} d \tau_2
\int\limits_0^{\hbar\beta} d \tau_3
\langle V^{x_0}_{{\rm int}} ( x ( \tau_1 ) )
V^{x_0}_{{\rm int}} ( x ( \tau_2 ) ) V^{x_0}_{{\rm int}} ( x ( \tau_3 ) )
\rangle^{x_0}_{\Omega,c} + \ldots \, . \hspace*{0.5cm}
\label{VPE}
\end{eqnarray}
The first term on the right-hand side is
the free energy of the local harmonic partition function
\begin{equation}
F^{x_0}_ \Omega\equiv
 -\frac{1}{\beta} \, \log  Z^{x_0}_{\Omega}
= \frac{1}{\beta} \, \log
\frac{\sinh \hbar \beta \Omega ( x_0 )/2}{\hbar \beta \Omega ( x_0)/2} \, .
\label{LFEN@}\end{equation}
The second term
contains the local harmonic expectation value
of the potential for which there exists the
above mentioned smearing formula
which we want to extend in this work.
The cumulant in the third term
is given by
the following combination of expectation values:
\begin{eqnarray}
\langle V^{x_0}_{{\rm int}} ( x ( \tau_1 ) )
V^{x_0}_{{\rm int}} ( x ( \tau_2 ) ) \rangle^{x_0}_{\Omega,c} =
\langle V^{x_0}_{{\rm int}} ( x ( \tau_1 ) )
V^{x_0}_{{\rm int}} ( x ( \tau_2 ) ) \rangle^{x_0}_{\Omega} -
\langle V^{x_0}_{{\rm int}} ( x ( \tau_1 ) ) \rangle^{x_0}_{\Omega}
\, \langle V^{x_0}_{{\rm int}} ( x ( \tau_2 ) ) \rangle^{x_0}_{\Omega}
\, . \label{C2}
\end{eqnarray}

By construction, the effective classical potential
$V^{{\rm eff,cl}} ( x_0 )$ in (\ref{VPE})
does not depend on the choice of the
frequency $\Omega ( x_0 )$ in the trial action (\ref{TA}). However,
when truncating the infinite sum (\ref{VPE}) after the $N$th order,
we obtain an approximation $W_N^{\Omega} ( x_0 )$ for the effective
classical potential $V^{{\rm eff, cl}} ( x_0 )$ with
an $\Omega ( x_0 )$-dependence, which decreases with increasing order
$N$ of the expansion.
With the expectation that
the optimal truncated sum $W_N^{\Omega} ( x_0 )$
depends minimally on the frequency $ \Omega(x_0)$, we
therefore determine $ \Omega(x_0)$ from
the extremality condition
\begin{equation}
\label{CON1}
\frac{\partial W_N^{\Omega} ( x_0 )}{\partial \Omega ( x_0 )} = 0 \, .
\end{equation}
If this has no solution,
we demand as the next-best condition
of minimal dependence on  $ \Omega(x_0)$
\cite{Kleinert3,Meyer,Kleinert2}
\begin{equation}
\label{CON2}
\frac{\partial^2 W_N^{\Omega} ( x_0 )}{\partial \Omega^2 ( x_0 )} = 0 \, .
\end{equation}
The result is called
the {\em optimal frequency\/}  $\Omega_N ( x_0 )$ of order $N$.
 It yields
the
truncated sum $W_N(x_0)\equiv W_N^{\Omega_N ( x_0 )} ( x_0 )$ which represents
the desired $N$th order approximation to the effective classical
potential $V^{{\rm eff, cl}} ( x_0 )$.
The first-order approximation
$W_1(x_0)$
coincides with
the original variational result
of Feynman and Kleinert \cite{Feynman2}
which satifies the Jensen-Peierls inequality
and guarantees the existence of an extremum (\ref{CON1}). \\[-3mm]

The accuracy of the approximate effective classical potential
$W_N ( x_0 )$ can be assessed by the following considerations
\cite{Kleinert2}. In the limit of high temperatures where $\beta \rightarrow 0$,
the approximation $W_N ( x_0 )$ becomes
exact for any $N$:
\begin{eqnarray}
\lim_{\beta \rightarrow 0} W_N ( x_0 ) = V ( x_0 ) \, .
\end{eqnarray}
At low temperatures, where $\beta \rightarrow \infty$, we
obtain from (\ref{PF2}) and (\ref{ECP}) an approximation
to the free energy in form of an integral
over the time-averaged position $x_0$
\begin{eqnarray}
F_N = - \frac{1}{\beta} \log  \left\{
\, \int\limits_{- \infty}^{+ \infty}\tilde d x_0 \exp \left[ - \beta
W_N ( x_0 ) \right] \right\} \, ,
\end{eqnarray}
whose integrand
is centered sharply  around the
minimum $x_N^{{\rm min}}$ of $W_N ( x_0 )$. Performing this integral
in the saddle point approximation yields an $N$-th order
approximation
$E_N^{(0)}$ for the
ground state energy $E^{(0)}$ of the quantum system.
\begin{eqnarray}
\label{SADDLE}
E_N^{(0)}= \min_{x_0}\lim_{\beta \rightarrow \infty}
W_N ( x_0 ) \, .
\end{eqnarray}
The same approximation
to the ground state energy
can also be obtained by
a variational resummation
 \cite{Kleinert2} of the
Rayleigh-Schr\"odinger perturbation series for $E^{(0)}$.
This will be shown in Section~V
for
the ground state energy of the Coulomb
potential up to the order $N=2$.
\section{Evaluation of Path Integrals}

In order to calculate
the different terms in the variational perturbation expansion (\ref{VPE}),
we must find efficient formulas
for evaluating local correlation functions
of the type (\ref{HEV}).
For this we observe
that, by fixing of
the temporal average at $\bar x=x_0$
in the path integral in (\ref{VPE}),
the zero Matsubara frequency $ \omega_0=0$ is removed from
the Fourier decomposition
of the periodic paths
\begin{equation}
\label{FS}
x ( \tau ) = x_0 + \sum_{m = 1}^{\infty} \left( x_m e^{i \omega_m \tau}
+ x_m^* e^{- i \omega_m \tau} \right), \,     ~~~
\omega_m = 2 \pi m / \hbar \beta \, .
 \end{equation}
In fact, the restricted integration measure $\oint {\cal D}
x ( \tau ) \tilde\delta (
\overline{x} - x_0 )$ in (\ref{HEV}) may be decomposed
into a product of ordinary integrals over real and imaginary parts
$x^{{\rm re}}_m$ and $x^{{\rm im}}_m$ of
the Fourier components $x_m$ according to
\cite{Kleinert2}:
\begin{equation}
\label{IND}
\oint {\cal D} x ( \tau ) \tilde \delta (
\overline{x} - x_0 ) =
\prod_{m = 1}^{\infty} \left(
\, \int\limits_{- \infty}^{+ \infty} d x^{{\rm re}}_m
\int\limits_{- \infty}^{+ \infty} d x^{{\rm im}}_m
\frac{\beta M \omega_m^2}{\pi} \right) \, .
\end{equation}
The zero-frequency component $x_0$ remains unintegrated.
With this decomposition, the basic local pair
correlation function
of the fluctuations
$ \delta x(\tau )$ in (\ref{devia})
can immediately be calculated
from (\ref{HEV}) as
a Matsubara sum
 without the zero mode:
\begin{eqnarray}
G_{\Omega}^{x_0}(\tau ,\tau ')&\equiv& \langle  \delta  x(\tau )
 \delta x(\tau' ) \rangle^{x_0}_{\Omega} =
\frac{2}{M \beta}
\sum _{m=1}^\infty
\frac{\cos \omega_m(\tau -\tau ')}{ \omega_m^2+ \Omega^2(x_0)} \, .
\label{GREEN0}\end{eqnarray}
Performing the sum yields the explicit result
\begin{eqnarray}
G_{\Omega}^{x_0}(\tau ,\tau ')=
\frac{\hbar}{2 M \Omega ( x_0 )} \, \left\{
\frac{\cosh \left[ \Omega ( x_0 ) | \tau -
\tau' | - \hbar \beta \Omega ( x_0 ) / 2 \right] }{\sinh
[ \hbar \beta \Omega ( x_0 ) / 2 ]}
- \frac{2}{\hbar \beta \Omega ( x_0 )} \right\} \, .
\label{GREEN}\end{eqnarray}
The first term
is  the ordinary oscillator correlation function
of frequency $\Omega $
\begin{equation}
\label{@D}
G_\Omega ( \tau , \tau' )\equiv\langle x(\tau )x(\tau' )\rangle_{\Omega}
 =
\frac{1}{M \beta}
\sum _{m=-\infty}^\infty
\frac{\cos \omega_m(\tau -\tau ')}{ \omega_m^2+ \Omega^2(x_0)} \, ,
\end{equation}
while the last term subtracts
the zero mode which is
 absent
in (\ref{GREEN0}).
This absence has
the important consequence
that
\begin{eqnarray}
\int\limits_{0}^{\hbar\beta} d \tau  \,
G_{\Omega}^{x_0} ( \tau , \tau' ) = 0 \, .
\end{eqnarray}

Using (\ref{GREEN}),
the
expectation values
in (\ref{VPE})
can easily be calculated for a polynomial potential
using
Wick's contraction
rules, by which the expectation values
can
be
reduced to
sums over products of pair correlation functions
$G_{\Omega}^{x_0}(\tau ,\tau ')$.
In order to
abbreviate the notation and to
emphasize the dimension (length)$^2$ of these quantities, we shall
denotes the local Green functions
in (\ref{GREEN}) from now on
by $a^2_{\tau \tau '}(x_0)$.
The harmonic expectation value of any odd power $n$ in the fluctuation
variable $ \delta x ( \tau )
$ is zero. For even
$n$,
the Wick expansion reads
\begin{eqnarray}
\label{WICK}
\left\langle \prod_{k = 1}^n  \delta x ( \tau_k )
\right\rangle^{x_0}_{\Omega} = \sum_P a^2_{\tau_{P ( 1 )} {\tau_{P ( 2 )} }}
( x_0 ) \cdots
a^2_{\tau_{P ( n - 1 )} {\tau_{P ( n )} }}
( x_0 ) \, ,
\end{eqnarray}
where the sum runs over all $( n - 1 )!!$ pair contractions.
For an exponential,
Wick's rule implies
\begin{eqnarray}
\left\langle \exp \left[ i \int\limits_0^{\hbar\beta} d \tau
j(\tau ) \delta x ( \tau )  \right]
\right\rangle^{x_0}_{\Omega} = \exp\left[ -\frac{1}{2}
\int\limits_0^{\hbar\beta} d\tau
\int\limits_0^{\hbar\beta} d\tau '
j ( \tau ) a^2_{\tau\tau'}(x_0)
j( \tau' )                \right] .
\label{@F}\end{eqnarray}
In the special case $j(\tau )=\sum _{k = 1}^n u_k \delta(\tau -\tau_k)$,
we obtain the important formula
for the expecation value
of a product of exponentials
\begin{eqnarray}
\left\langle \prod_{k=1}^ne^{ i u_k  \delta x ( \tau_k )}
\right\rangle^{x_0}_{\Omega} = \exp \left[ - \frac{1}{2}
\sum_{k = 1}^n \sum_{k' = 1}^n u_k a^2_{\tau_k \tau_{k'}} ( x_0 )
u_{k'} \right]
\, .
\label{@expWick}\end{eqnarray}
After Fourier-decomposing the functions $F_1(x), \ldots , F_n(x)$
in (\ref{HEV}), formula (\ref{@expWick})
yields directly the desired smearing formula
\cite{Kuerzinger}
\begin{eqnarray}
&&\!\!\!\!\!\!\!\!\!\!\!\!\!\!\!\!\left\langle
F_1 ( x ( \tau_1 ) ) \cdots F_n ( x ( \tau_n ) )
\right\rangle^{x_0}_{\Omega} = \left[ \prod_{k = 1}^n \,
\int\limits_{- \infty}^{+ \infty}
d x_k F_k ( x_k ) \right]\hspace*{2cm} \nonumber \\
&&~~~~~~\times \, \frac{1}{\sqrt{ ( 2 \pi )^n \mbox{Det} \left[
a^2_{\tau_k \tau_{k'}} ( x_0 )
\right]}}
\exp \left[ - \frac{1}{2} \sum_{k = 1}^n \sum_{k' = 1}^n
  \delta x_k \,a^{-2}_{\tau_k \tau_{k'}}( x_0 )
   \,\delta x_{k'}  \right] \, ,
\label{SM}
\end{eqnarray}
where $a^{-2}_{\tau_k \tau_{k'}}
( x_0 )$ denotes the inverse of the symmetric $n\times n$-matrix
$a^2_{\tau_k \tau_{k'}}
( x_0 )$.
This smearing formula determines
the different harmonic expectation values in the
variational perturbation expansion (\ref{VPE})
as convolutions with Gaussian functions.\\[-3mm]

For $n = 1$, the smearing formula (\ref{SM}) reduces
to the previous one \cite{Feynman2,Kleinert2}
\begin{equation}
\label{SP}
\langle F_1 ( x ( \tau_1 ) ) \rangle^{x_0}_{\Omega} =
\int\limits_{- \infty}^{+ \infty} d x_1 F_1 ( x_1 )
\frac{1}{\sqrt{2 \pi a^2 ( x_0 )}} \exp \left[ -
\frac{( x_1 - x_0 )^2}{2 a^2 ( x_0 )} \right] \,,
\end{equation}
where $a^2(x_0)$ denotes the $\tau $-independent
diagonal matrix element  $a^2_{\tau \tau }(x_0)$.
For polynomials $F_1(x)$, the smearing
formula
(\ref{SM}) reproduces Wick's rule: Odd powers in $ \delta x ( \tau ) $
have  vanishing
local correlation functions,
whereas even powers result in (\ref{WICK}), which for coinciding times $\tau _k$
reduces to
\begin{eqnarray}
\label{@asquare}
\left\langle  \left[\delta x ( \tau_k ) \right]^n
\right\rangle^{x_0}_{\Omega}= ( n - 1 )!! \,\,a^{n} ( x_0 ) \, .
\end{eqnarray}
For two functions $F_1(x)$ and $F_2(x)$,
our smearing formula (\ref{SM}) reads, more explicitly,
\begin{eqnarray}
\,\left\langle F_1 ( x ( \tau_1 ) )
F_2 ( x ( \tau_2 ) ) \right\rangle^{x_0}_{\Omega}
= \int\limits_{- \infty}^{+ \infty} d x_1 \int\limits_{- \infty}^{+ \infty}
d x_2 \, F_1 ( x_1 ) F_2 ( x_2 ) \,\frac{1}{\sqrt{( 2\pi )^2 [
a^4 ( x_0 ) - a_{\tau_1 \tau_2}^4 ( x_0 ) ]}}
~~~~~~~~~ \nonumber \\
 \times \exp \left\{ - \frac{a^2(x_0) ( x_1 - x_0 )^2 - 2 a^2_{\tau_1 \tau_2}
( x_0 ) ( x_1 - x_0 ) ( x_2 - x_0 ) + a^2 ( x_0 ) ( x_2 -
x_0 )^2}{2 [ a^4 ( x_0 ) - a_{\tau_1 \tau_2}^4 ( x_0 )]}
\right\} \, . \label{SPE}
\end{eqnarray}
Specializing $F_2(x)$ to the square of the function $\delta x$,
we obtain the useful rule
\begin{eqnarray}
\left\langle F_1 ( x ( \tau_1 ) ) \left[ \delta x ( \tau_2 ) \right] ^2
\right\rangle^{x_0}_{\Omega} & = &
a^2 ( x_0 )
\left[ 1 - \frac{a_{\tau_1 \tau_2}^4
( x_0 )}{a^4 ( x_0 )} \right]
\left\langle F_1 ( x( \tau_1 ) ) \right\rangle^{x_0}_{\Omega}
\nonumber \\
&+&
\frac{a^4_{\tau_1 \tau_2}
( x_0 )}{a^4( x_0 )}
\left\langle F_1 ( x ( \tau_1 ) ) \left[ \delta x ( \tau_1 ) \right]^2
\right\rangle^{x_0}_{\Omega}
\, , \label{EXX1}
\end{eqnarray}
which reduces the smearing procedure for different times
$\tau_1$ and $\tau_2$ to corresponding ones at equal times $\tau_1 = \tau_2$.
With this we immediately yield, for instance,
\begin{eqnarray}
 \label{EXX2}
\left\langle\left[ \delta x ( \tau_1 )  \right]^2
\left[ \delta x ( \tau_2 )  \right]^2 \right\rangle^{x_0}_{\Omega}
= a^4 ( x_0 ) + 2 a^4_{\tau_1 \tau_2} ( x_0 ) \, .
\end{eqnarray}

In three dimensions,
the trial potential
contains a
$3\times 3$-frequency matrix $\Omega_{ij}$ depending on the
time-averaged position ${\bf x}_0$
and reads
\begin{eqnarray}
\frac{M}{2} \sum_{i,j = 1}^3 \Omega^2_{ij} ( {\bf x}_0 ) ( x_i - x_{0i} )
( x_j - x_{0j} ) \, ,
\end{eqnarray}
while the interaction potential
 (\ref{@intpot})  becomes
\begin{equation}
V^{{\bf x}_0}_{{\rm int}} ( {\bf x} ) = V({\bf x})
- \frac{M}{2}  \sum_{i,j = 1}^3 \Omega^2_{ij} ( {\bf x}_0 ) ( x_i - x_{0i} )
( x_j - x_{0j} ) \, .
\label{@intpot2g}\end{equation}
Assuming the potential to depend only on $r = | {\bf x} |$, i.e.
$V({\bf x})=v(r)$, the frequency matrix
possesses only two invariant matrix elements,
a longitudinal one $\Omega_L ( r_0 )$ and a transversal
one $\Omega_T ( r_0 )$ \cite{Kleinert2,Janke}.
The interaction potential (\ref{@intpot2g})
can then be decomposed
into a longitudinal and a transversal part according to
\begin{eqnarray}
\Omega^2_{ij} ( {\bf x}_0 ) = \Omega^2_L( r_0 ) \,
\frac{x_{0i} x_{0j}}{r_0^2}
+ \Omega^2_T( r_0 ) \, \left( \delta_{ij} -
\frac{x_{0i} x_{0j}}{r_0^2} \right) \, ,
\end{eqnarray}
so that (\ref{@intpot2g})
may be rewritten
as
\begin{equation}
V^{{\bf x}_0}_{{\rm int}} ( {\bf x} ) = v(r)
- \frac{M}{2} \left\{
\Omega^2_{L}  ( r_0 )\left[  \delta {\bf x}\right]_L^2
+\Omega^2_{T}  ( r_0 )\left[  \delta {\bf x}\right]_T^2\right\} ,
\label{@intpot2p}\end{equation}
with obvious definitions
of the longitudinal and transverse projections
$  \delta {\bf x}_L $ and
$  \delta {\bf x}_T $ of the fluctuations $ \delta{\bf x}$.

To first order,
the anisotropic generalization of the smearing formula
(\ref{SP}) reads
 \cite{Kleinert2,Janke}
\begin{eqnarray}
\left\langle F_1 ( {\bf x} ( \tau_1 ) \right\rangle^{r_0}_{\Omega_T,
\Omega_L} = \int\limits_{- \infty}^{+ \infty} d^3 x_1 F_1 ( {\bf x}_1 )
\, \frac{1}{\sqrt{( 2 \pi )^3 a_T^4 a_L^2}} \exp \left\{
- \frac{{\bf x}_{1T}^2}{2 a_T^2} - \frac{(x_{1L} - r_0 )^2}{2 a_L^2} \right\}
\, .
\end{eqnarray}
For the squares of transverse and longitudinal fluctuations,
this generalizes (\ref{@asquare}) with $n = 2$ to
\begin{eqnarray}
\left\langle \left[ \delta {\bf x} ( \tau_1 )  \right]^2_T
\right\rangle^{r_0}_{\Omega_T,\Omega_L} & = & 2 a_T^2 \, ,
~~~~~~~~~\left\langle \left[ \delta {\bf x} ( \tau_1 ) \right]^2_L
\right\rangle^{r_0}_{\Omega_T,\Omega_L} ~~= ~~ a_L^2 \, .
 \label{EX2}
\end{eqnarray}

The second-order
smearing formula (\ref{SPE})
becomes in
three dimensions
\begin{eqnarray}
&&\!\left\langle F_1 ( {\bf x} ( \tau_1 ) F_2 ( {\bf x} ( \tau_2 )
\right\rangle^{r_0}_{\Omega_T,\Omega_L} = \int\limits_{- \infty}^{+ \infty}
d^3 x_1 \int\limits_{- \infty}^{+ \infty} d^3 x_2 \, F_1 ( {\bf x}_1 )
F_2 ( {\bf x}_2 ) \hspace*{2cm}  \nonumber \\
&&~~~~~\times \, \frac{1}{( 2 \pi )^3 ( a^4_T - a^4_{T \tau_1 \tau_2} )
\sqrt{a_L^4 - a^4_{L \tau_1 \tau_2} }}
\exp \left\{ - \frac{a_T^2 {\bf x}_{1T}^2 - 2 a^2_{T \tau_1 \tau_2}
{\bf x}_{1T}
{\bf x}_{2T} + a^2_T {\bf x}_{2T}^2 }{2 ( a^4_T - a^4_{T \tau_1
\tau_2} )} \right\}
\nonumber \\
&&~~~~~\times \, \exp \left\{
- \frac{a_L^2 ( x_{1L} - r_0 )^2 - 2 a^2_{L \tau_1 \tau_2} ( x_{1L} - r_0 )
( x_{2L} - r_0 ) + a_L^2 ( x_{2L} - r_0 )^2 }{2 ( a^4_L - a^4_{L
\tau_1 \tau_2} )}
\right\} \, , \label{ANISO}
\end{eqnarray}
so that the rule (\ref{EXX1})
for expectation values is generalized to
\begin{eqnarray}
\left\langle F_1 ( {\bf x} ( \tau_1 )  \left[ \delta {\bf x} ( \tau_2 )
 \right]^2_T
\right\rangle^{r_0}_{\Omega_T,\Omega_L} & = & 2 a^2_T \left[ 1 -
\frac{a^4_{T \tau_1 \tau_2}}{a^4_T} \right]
\left\langle F_1 ( {\bf x} ( \tau_1 ) )
\right\rangle^{r_0}_{\Omega_T,\Omega_L} \nonumber \\
\label{RULE1}
&&+ \frac{a^4_{T \tau_1 \tau_2}}{a^4_T}
\left\langle F_1 ( {\bf x}  ( \tau_1 ) )
\left[ \delta {\bf x} ( \tau_1 )  \right]^2_T
\right\rangle^{r_0}_{\Omega_T,\Omega_L} \, , \\
\left\langle F_1 ( {\bf x} ( \tau_1 )  \left[ \delta {\bf x} ( \tau_2 )
 \right]^2_L
\right\rangle^{r_0}_{\Omega_T,\Omega_L} & = &        ~
a_L^2 \left[ 1 - \frac{a^4_{L \tau_1 \tau_2}}{a^4_L} \right]
\left\langle F_1 ( {\bf x} ( \tau_1 ) )
\right\rangle^{r_0}_{\Omega_T,\Omega_L} \nonumber \\
& & + \frac{a^4_{L \tau_1 \tau_2}}{a^4_L}
\left\langle F_1 ( {\bf x} ( \tau_1 )
[ \delta {\bf x} ( \tau_1 ) ]^2_L
\right\rangle^{r_0}_{\Omega_T,\Omega_L} \, .
\label{RULE2}
\end{eqnarray}
Specializing $F_1 ( {\bf x} ) $ to a quadratic function, we obtain
the corresponding generalizations of (\ref{EXX2})
\begin{eqnarray}
\left\langle \left[ \delta {\bf x} ( \tau_1 ) \right]^2_T
\left[  \delta{\bf x} ( \tau_2 )  \right]^2_T
\right\rangle^{r_0}_{\Omega_T,\Omega_L}
 & = & 4 a^4_T + 4 a^4_{T \tau_1 \tau_2} \, , \\
\left\langle \left[ \delta {\bf x} ( \tau_1 ) \right]^2_T
\left[ \delta {\bf x} ( \tau_2 ) \right]^2_L
\right\rangle^{r_0}_{\Omega_T,\Omega_L}
 & = & 2 a^2_T a^2_L \, , \\
\left\langle \left[ \delta {\bf x} ( \tau_1 )  \right]^2_L
\left[  \delta{\bf x} ( \tau_2 )  \right]^2_L
\right\rangle^{r_0}_{\Omega_T,\Omega_L}
 & = & a^4_L + 2 a^4_{L \tau_1 \tau_2} \, .
\end{eqnarray}

\section{Application to Coulomb Potential}

Let us demonstrate the use of the new
smearing formulas
by
calculating the
effective classical potential of the three-dimensional
Coulomb potential
\begin{eqnarray}
\label{C}
v ( r ) = - \frac{e^2}{r}
\end{eqnarray}
up to the second order in the variational perturbation expansion, thus
going beyond
the known first-order results in \cite{Kleinert2,Janke}.
For harmonic and Coulomb potentials
we
express
 the Coulomb
potential (\ref{C})
as a   ``proper-time"
integral
\begin{eqnarray}
\frac{1}{r} = 4 \pi \int\limits_{- \infty}^{+ \infty}
\frac{d^3 k}{( 2 \pi)^3} \int\limits^{+ \infty}_0 d \sigma e^{
 - \sigma {\bf k}^2
- i {\bf k} {\bf x}} \, ,
\label{@prop}\end{eqnarray}
where $ \sigma$ has the dimension (length)$^2$, and
find the expectation value
\begin{eqnarray}
\left\langle \frac{1}{| {\bf x} ( \tau_1 )|}
\right\rangle^{r_0}_{\Omega_T,\Omega_L}
& = & \sqrt{\frac{2 a_L^2}{\pi}} \,\int\limits_0^1 d \lambda \,
\frac{1}{( a^2_T - a^2_L ) \lambda^2 + a^2_L} \,
\exp \left\{ - \frac{r_0^2}{2 a_L^2} \lambda^2 \right\} \, .
\end{eqnarray}
From a straight-forward
three-dimensional extension of (\ref{VPE}),
the first-order variational approximation
to the effective classical potential is then
\cite{Kleinert2,Janke}
\begin{eqnarray}
W_1^{\Omega_T,\Omega_L} ( r_0 ) & = & \frac{2}{\beta} \log
\frac{\sinh [ \hbar \beta \Omega_T / 2 ]}{\hbar \beta \Omega_T / 2}
+ \frac{1}{\beta} \log
\frac{\sinh [ \hbar \beta \Omega_L / 2 ]}{\hbar \beta \Omega_L / 2}
- \frac{M}{2} \left\{ 2 \Omega_T^2 a_T^2 + \Omega^2_L a_L^2 \right\}
\nonumber \\
& & \label{W1}
- e^2 \sqrt{\frac{2 a_L^2}{\pi}} \,\int\limits_0^1 d \lambda \,
\frac{1}{ ( a^2_T - a^2_L ) \lambda^2 + a^2_L } \,
\exp \left\{ - \frac{r_0^2}{2 a_L^2} \lambda^2 \right\} \, ,
\label{@firstorderW}
\end{eqnarray}
where we have ommitted the argument $r_0$ from all functions on the
right-hand side.
A similar
expression was derived
in the isotropic approximation $\Omega_L = \Omega_T$
with the help of
Gaussian wave packets
in the context of plasma physics
for the purpose of faster
molecular dynamics simulations
\cite{Bachmann2,Klakow,Kelbg}.\\[-3mm]

By inserting the
``proper-time"
integral
for the Coulomb potential
(\ref{@prop})
into the rules
(\ref{RULE1}) and (\ref{RULE2}),
we find
\begin{eqnarray}
\left\langle \frac{1}{| {\bf x} ( \tau_1 ) |}
\left[ \delta {\bf x} ( \tau_2 ) \right]^2_T
\right\rangle^{r_0}_{\Omega_T,\Omega_L} & = &
\sqrt{\frac{2 a_L^2}{\pi}} \int\limits_0^1 d \lambda \,
\exp \left\{ - \frac{r^2_0}{2 a^2_L} \lambda^2 \right\}
\nonumber \\
& & \times \left\{
\frac{2 a^2_T}{ ( a^2_T - a^2_L ) \lambda^2 + a^2_L }
- \frac{2 a^4_{T \tau_1 \tau_2} \lambda^2}{[ ( a^2_T - a^2_L )
\lambda^2 + a^2_L ]^2} \right\} \, ,
\label{APP1} \\
\left\langle \frac{1}{| {\bf x} ( \tau_1 ) |}
\left[ \delta {\bf x} ( \tau_2 ) \right]^2_L
\right\rangle^{r_0}_{\Omega_T,\Omega_L} & = &
\sqrt{\frac{2 a_L^2}{\pi}} \int\limits_0^1 d \lambda \,
\exp \left\{ - \frac{r^2_0}{2 a^2_L} \lambda^2 \right\} \,
\frac{a^6_L + a^4_{L \tau_1 \tau_2}
[ r^2_0 \lambda^4 - a^2_L \lambda^2]}{a^4_L [
( a^2_T - a^2_L ) \lambda^2 + a^2_L ]} \, .
\label{APP2}
\end{eqnarray}
Note that these results
are also special cases of the general
expectation value
\begin{eqnarray}
\left\langle  \frac{1}{| {\bf x} ( \tau_1 ) |}
F ( {\bf x}( \tau_2 ) )\right\rangle^{r_0}_{\Omega_T,\Omega_L} = \frac{1}{2 \pi^2}
\int\limits_0^{+ \infty} d \sigma \,
\frac{{\displaystyle \exp \left\{ -
\frac{a^2_L r^2_0}{2 [ a^4_L - a^4_{L \tau_1 \tau_2}
+ 2 a^2_L \sigma]} \right\} } }{[
a^4_T - a^4_{T \tau_1 \tau_2} + 2 a^2_T \sigma ] \sqrt{a^4_L -
a^4_{L \tau_1 \tau_2} + 2 a^2_L \sigma} }
\int\limits_{- \infty}^{+ \infty} d^3 x F ( {\bf x} ) \nonumber \\
\times \exp \left\{ -
\frac{(a^2_T + 2 \sigma) {\bf x}^2_{T}}{2
[a^4_T - a^4_{T \tau_1 \tau_2} + 2 a^2_T \sigma]}
- \frac{( a^2_L + 2 \sigma ) ( x_{L} - r_0 )^2 + 2 a^2_{L \tau_1 \tau_2} r_0
( x_{L} - r_0 )}{2 [a^4_L - a^4_{L \tau_1 \tau_2} + 2 a^2_L \sigma]}
\right\} \, ,
\end{eqnarray}
which gives us the additional local expectation value

\begin{eqnarray}
\label{CCE}
&&\!\!\!\!\!
\left\langle \frac{1}{| {\bf x} ( \tau_1 )|} \frac{1}{| {\bf x} ( \tau_2 )|}
\right\rangle^{r_0}_{\Omega_T,\Omega_L} =
\frac{2}{\pi} \int\limits_0^{+ \infty}
d \sigma_1 \int\limits_0^{+ \infty} d \sigma_2 \,
\frac{1}{[ a^2_T + 2 \sigma_1 ]
[ a^2_T + 2 \sigma_2 ] - a^4_{T \tau_1 \tau_2} } \nonumber\\
&&~~~~\times \frac{1}{\sqrt{[ a^2_L + 2 \sigma_1 ] [ a^2_L + 2 \sigma_2 ]
- a^4_{L \tau_1 \tau_2} } }
\exp \left\{ -
\frac{r^2_0 [a^2_L + \sigma_1 + \sigma_2 - a^2_{L \tau_1 \tau_2}]}{[
a^2_L + 2 \sigma_1 ] [ a^2_L + 2 \sigma_2 ] -
a_{L \tau_1 \tau_2}^4 } \right\} \, .
\end{eqnarray}
From these smearing results
we calculate
the
connected correlation functions
of the interaction potential  (\ref{@intpot2g})
according to the cumulant law
(\ref{C2}),
and insert these into (\ref{VPE})
to obtain
 the second-order
     effective classical potential
\begin{eqnarray}
W_2^{\Omega_T,\Omega_L} ( r_0 ) & = & W_1^{\Omega_T,\Omega_L} ( r_0 )
\nonumber \\
& & +
\frac{e^2 M}{2 \hbar} \sqrt{\frac{2 a^2_L}{\pi}} \int\limits_0^1 d \lambda
\left\{ \frac{2 \Omega_T l^4_T \lambda^2}{[ ( a^2_T - a^2_L )
\lambda^2 + a^2_L ]^2} - \frac{\Omega_L l^4_L [ r^2_0 \lambda^4 -
a^2_L \lambda^2 ] }{a^4_L [ ( a^2_T - a^2_L )
\lambda^2 + a^2_L ] } \right\}
\exp \left\{ - \frac{r^2_0}{2 a^2_L} \lambda^2 \right\} \nonumber \\
& & - \frac{M^2 [ 2 \Omega^3_T l^4_T + \Omega^3_L l^4_L ]}{4 \hbar}
- \frac{e^4}{2\hbar^2 \beta}\int\limits_0^{\hbar \beta} d \tau_1
\int\limits_0^{\hbar \beta} d \tau_2
 \left\langle \frac{1}{| {\bf x} ( \tau_1 )|} \frac{1}{| {\bf x} ( \tau_2 )|}
\right\rangle^{r_0}_{\Omega_T,\Omega_L,c}  \label{W2}
\end{eqnarray}
with the abbreviation
\begin{eqnarray}
\label{AAB}
l^4_{T,L} = \frac{\hbar
\left[4 + \hbar^2 \beta^2 \Omega^2_{T,L}
- 4 \cosh \hbar \beta \Omega_{T,L} + \hbar \beta
\Omega_{T,L}  \sinh \hbar \beta \Omega_{T,L} \right]}{8
\beta M^2 \Omega_{T,L}^3  \sinh^2 [ \hbar \beta
\Omega_{T,L} / 2 ]} \, ,
\end{eqnarray}
which is a function of $r_0$  of dimension (length)$^4$.
After extremizing (\ref{W1}) and (\ref{W2}) with respect to the
trial frequencies $\Omega_T, \Omega_L$ according to (\ref{CON1}),
which has to be done numerically
we obtain
the first- and second-order approximations for the
effective
classical potential of the Coulomb system.
The isotropic approximations $\Omega_T = \Omega_L$ in the first and
second order are
plotted in Fig.~1
for various temperatures.
The second-order curves lie all
below
the first-order ones, and the difference between the two
decreases
with increasing temperature and increasing distance from the
origin. Figure~2 shows exemplarily that the anisotropic ap\-pro\-ximation
slightly deviates from the isotropic one. The difference between both
is only visible for intermediate distances from the origin.\\

\section{Zero-Temperature Limit}

In order to check our results we
take (\ref{W1}) and (\ref{W2}) to the limit $T \rightarrow 0$, where
$W_1^{\Omega} ( r_0 )$ and $W_2^{\Omega} ( r_0 )$ reduce at $r_0 = 0$,
according to (\ref{SADDLE}), to the
ground state energy of the Coulomb system.
At $r_0 = 0$, the frequencies are isotropic $\Omega_L = \Omega_T = \Omega$
for symmetry reasons, thus simplifying the evaluation
(\ref{W1}) and (\ref{W2}).
Taking into account the low
temperature limit of the
two-point correlations
(\ref{GREEN})
\begin{eqnarray}
\label{LOA}
\lim_{\beta \rightarrow \infty} a^2_{\tau_k \tau_{k'}} (
{\bf x}_0 ) = \frac{\hbar}{2 M \Omega ( {\bf x}_0 )}
e^{ - \Omega ( {\bf x}_0 ) \left| \tau_k - \tau_{k'} \right|} \, ,
\end{eqnarray}
we immediately deduce for the first order approximation (\ref{W1})
at ${\bf x}_0 = {\bf 0}$
with $\Omega = \Omega ( {\bf 0} )$ the limit

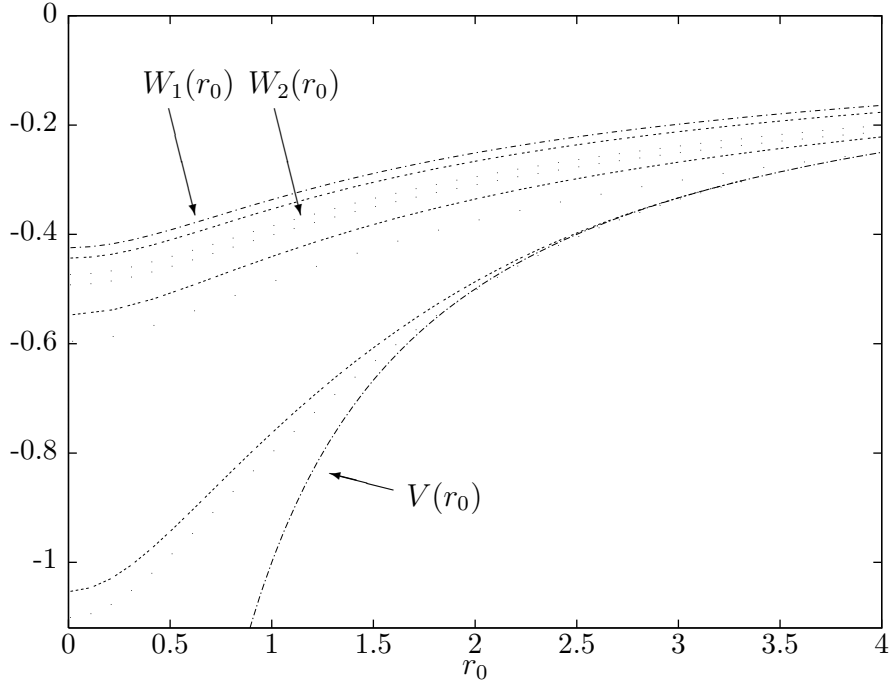
\begin{figure}[t]
~~~~\input mittel.tps
\begin{picture}(5,5)
\end{picture}
~\\~\\
\caption{Isotropic approximations to the effective classical potential of the
Coulomb system in the first (lines) and second order (dots). The
temperatures are 0, 0.1, 0.01 and $\infty$ from the top to the bottom
in atomic units. The high temperature limit is the same for all
approximations $W_N ( r_0 )$.}
\end{figure}
\hspace*{0.5cm}
\begin{eqnarray}
E_1^{(0)} ( \Omega ) = \lim_{\beta \rightarrow \infty}
W_1^{\Omega} ( {\bf 0} )
 = \frac{3}{4} \hbar \Omega - \frac{2}{\sqrt{\pi}}
\sqrt{\frac{M \Omega}{\hbar}} e^2 \, . \label{E1}
\end{eqnarray}

To second order, the limit is more involved.
Performing the integrals over
$\sigma_1$ and $\sigma_2$ in (\ref{CCE}), we obtain
with (\ref{C2})
the connected correlation function
\begin{eqnarray}
\label{LOC}
\left\langle \frac{1}{| {\bf x} ( \tau_1 )|} \frac{1}{| {\bf x} ( \tau_2 )|}
\right\rangle^{{\bf 0}}_{\Omega,c} = \frac{1}{a_{\tau_1 \tau_2}^4 ( {\bf 0} )}
- \frac{2}{\pi a^4_{\tau_1 \tau_2} ( {\bf 0} )}
\arctan \sqrt{\frac{a_{\tau_1 \tau_2}^2 ( {\bf 0} )}{a_{\tau_1 \tau_2}
( {\bf 0} )} - 1} - \frac{2}{\pi a^2_{\tau_1 \tau_2} ( {\bf 0} )} \, .
\end{eqnarray}
Inserting here the zero-temperature limit (\ref{LOA}),
we can integrate these expressions
over
the imaginary times
$\tau_1 , \tau_2 \in [ 0 , \hbar \beta ]$, and find for
large $\beta $
\begin{eqnarray}
\label{CCC}
\!\!\!\!\!\!\!\!\!\!\!\!\!\!\!\!\!\!\!\int\limits_0^{\hbar \beta} d \tau_1 \int\limits_0^{\hbar \beta} d \tau_2
\left\langle \frac{1}{| {\bf x} ( \tau_1 )|} \frac{1}{| {\bf x} ( \tau_2 )|}
\right\rangle^{{\bf 0}}_{\Omega,c} =
\frac{4 M}{\hbar \Omega} \left\{
e^{\hbar\beta\Omega} - 1 - \hbar \beta \Omega
- \frac{\hbar^2 \beta^2 \Omega^2}{\pi}
- \frac{2}{\pi} \right. \nonumber~~~~~~~~~~~~ \\
~~~~~~~~~\times \left.
\left[ e^{\hbar\beta\Omega} \arcsin \sqrt{1 - e^{- 2 \hbar \beta \Omega}}
+ \frac{1}{2} \log  \alpha ( \beta ) - \frac{1}{8} \left[ \log  \alpha
( \beta ) \right]^2 - \frac{1}{2} \int\limits_{\alpha ( \beta )}^1
d u \, \frac{\log  u}{1 + u} \right] \right\},
\end{eqnarray}
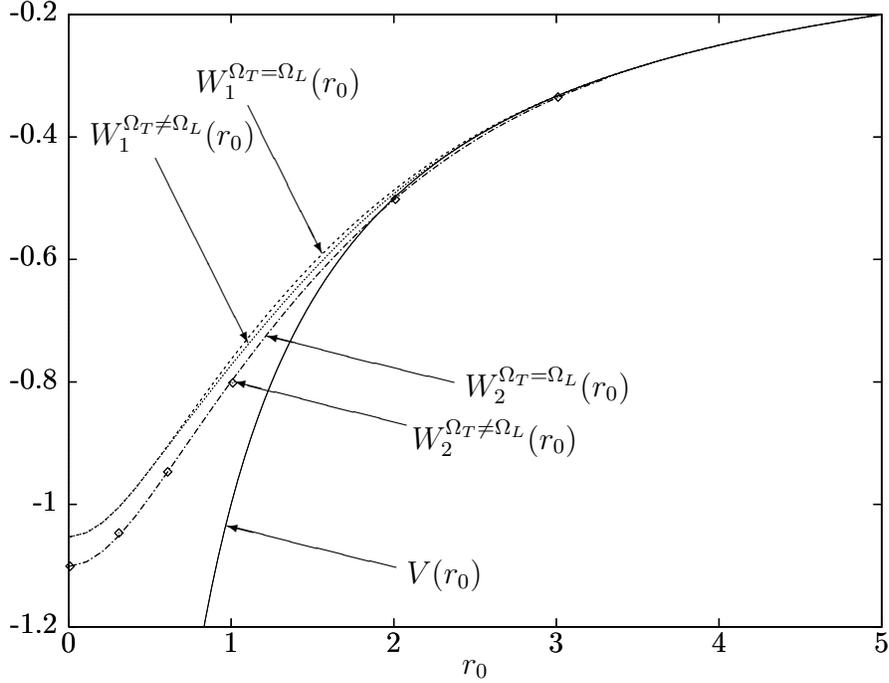
\begin{figure}[htbp]
\vspace*{-1cm}
~~~~\input ane.tps
\begin{picture}(5,5)
\end{picture}
~\\~\\
\caption{Isotropic and anisotropic approximations to the effective
classical potential of the
Coulomb system in the first and second order at the
temperature 0.1 in atomic units.
The lowest line represents the high temperature limit in which all
isotropic and anisotropic approximations coincide.}
\end{figure}
\vspace*{0.3cm}

with the abbreviation
\begin{eqnarray}
\label{AAA}
\alpha ( \beta ) = \frac{1 - \sqrt{1 - e^{- 2 \hbar \beta \Omega}}}{
1 + \sqrt{1 - e^{- 2 \hbar \beta \Omega}}} \, .
\end{eqnarray}
Inserting this into
(\ref{W2}) and going to the limit $ \beta\rightarrow \infty$,
we find the
ground state energy
\begin{eqnarray}
E_2^{(0)} ( \Omega ) =
\lim_{\beta \rightarrow \infty}
W_2^{\Omega} ( {\bf 0} ) =
\frac{9}{16} \hbar \Omega - \frac{3}{2 \sqrt{\pi}}
\sqrt{\frac{M \Omega}{\hbar}} e^2 - \frac{4}{\pi} \left( 1 + \log  2
- \frac{\pi}{2} \right) \frac{M}{\hbar^2} e^4 . \label{E2}
\end{eqnarray}
Postponing for a moment the extremization of (\ref{E1}) and (\ref{E2})
with respect to the trial frequency $\Omega$,
let us first rederive this result
from a variational treatment
of the ordinary Rayleigh-Schr\"odinger perturbation expansion
for the ground-state energy.
\section{Variational Treatment of Rayleigh-Schr\"odinger Perturbation Expansion}

According to the rules layed out in
\cite{Kleinert2},
we must first calculate the ground-state energy
for the Coulomb potential
in the presence of a harmonic potential of frequency $ \omega$:
\begin{eqnarray}
\label{V3}
V_{{\rm aux}}
( {\bf x} ) = \frac{M}{2} \omega^2 {\bf x}^2 - \frac{e^2}{| {\bf x} |} \, .
\end{eqnarray}
After this, we make the trivial replacement
$ \omega\rightarrow  \sqrt{ \Omega^2+ \omega^2- \Omega^2}$,
and reexpand  the energy in powers of
the difference $ \omega^2- \Omega^2$, considering
this quantity as being of the order $e^2$ and truncating the reexpansion accordingly.
At the end we set $\omega=0$, since the original Coulomb system
contains no oscillator potential.
Note that this limit is equivalent to a strong-coupling limit of (\ref{V3})
with respect to $e^2$.
The result of this treatment
will coincide precisely with
the expansions (\ref{E1}) and (\ref{E2}), respectively.\\[-3mm]

The  Rayleigh-Schr\"odinger  perturbation expansion of the ground state energy
$E_N^{{\rm aux}} ( \omega )$
for the potential (\ref{V3})
requires the knowledge of the matrix elements
of the Coulomb potential (\ref{C})
with respect to the eigenfunctions
$\psi_{n,l,m} ( r , \vartheta , \varphi )$
of the harmonic oscillator with the frequency $\omega$
\cite{Kleinert2}:
\begin{eqnarray}
\label{MATT}
V_{n,l,m;n',l',m'} = \int\limits_0^{2 \pi} d \varphi
\int\limits_0^{\pi} d \vartheta \sin \vartheta \int\limits_0^{\infty}
d r \, r^2 \, \psi^*_{n,l,m} ( r , \vartheta , \varphi )
\frac{- e^2}{r} \psi_{n',l',m'} ( r , \vartheta , \varphi )\, ,
\end{eqnarray}
\begin{eqnarray}
\psi_{n,l,m} ( r , \vartheta , \varphi ) & = &\sqrt{ \frac{2n!}{\Gamma
( n + l + 3/2 )}} \,\phantom{}^4\sqrt{\frac{M \omega}{\hbar}}
\,\left( \frac{M \omega}{\hbar} r^2 \right)^{(l + 1)/2}
\nonumber \\
&&\times L_n^{l + 1/2} \left( \frac{M \omega}{\hbar} r^2 \right)
\exp \left\{ - \frac{M \omega}{2 \hbar} r^2 \right\} Y_{l,m}
( \vartheta , \varphi ) \, .
\label{PSI}
 \end{eqnarray}
Here $n$ denotes the radial quantum number, $L^{\alpha}_n ( x )$ the
Laguerre polynomials \cite{Gradshteyn}, and $Y_{l,m} ( \vartheta , \varphi )$
the spherical harmonics obeying the orthonormality relation
\begin{eqnarray}
\label{ORT}
\int\limits_0^{2 \pi} d \varphi \int\limits_0^{\pi} d \vartheta
\sin \vartheta \, Y^*_{l,m} ( \vartheta , \varphi ) Y_{l',m'}
( \vartheta , \varphi ) = \delta_{l,l'} \delta_{m,m'} \, .
\end{eqnarray}
Inserting (\ref{PSI}) into (\ref{MATT}), and evaluating the integrals
with Eq. (2.19.14.15) in Ref. \cite{Prudnikov},
we find
\begin{eqnarray}
V_{n,l,m;n',l',m'} & = & - e^2 \, \sqrt{\frac{M \omega}{\pi \hbar}} \,
\frac{\Gamma ( l + 1 ) \Gamma ( n + 1/2 )}{\Gamma ( l + 3/2 )}
\,\sqrt{\frac{\Gamma ( n' + l + 3/2 )}{n! n'! \Gamma (n + l + 3/2 )}}
\nonumber \\
&&\times
\phantom{}_3 F_2 \left( - n' , l + 1 , \frac{1}{2} ; l + \frac{3}{2} ,
\frac{1}{2} - n ; 1 \right) \delta_{l,l'} \,\delta_{m,m'} \, ,
\label{MAA}
\end{eqnarray}
with the generalized hypergeometric series \cite{Gradshteyn}
\begin{eqnarray}
\phantom{}_3 F_2 ( \alpha_1 , \alpha_2 , \alpha_3 ; \beta_1 , \beta_2 ; x ) =
\sum_{k = 0}^{\infty} \frac{( \alpha_1 )_k ( \alpha_2 )_k
( \alpha_3 )_k}{( \beta_1 )_k ( \beta_2 )_k} \,\frac{x^k}{k!} \, ,
\end{eqnarray}
and the Pochhammer symbol $( \alpha )_k = \Gamma ( \alpha + k ) /
\Gamma ( \alpha )$.\\[-3mm]

These matrix elements are now inserted
into the
Rayleigh-Schr\"odinger perturbation expansion for the ground state
energy
\begin{eqnarray}
E_{{\rm aux}} ( \omega )& = &E_{0,0,0} + V_{0,0,0;0,0,0} +
\sum_{n,l,m} ' \frac{V_{0,0,0;n,l,m} V_{n,l,m;0,0,0}}{E_{0,0,0} -
E_{n,l,m}}
-  \sum_{n,l,m} '  V_{0,0,0;0,0,0}
\frac{V_{0,0,0;n,l,m} V_{n,l,m;0,0,0} }{\left[ E_{0,0,0} -
E_{n,l,m} \right]^2}
 \nonumber \\
\label{XO}
&&+ \sum_{n,l,m} ' \sum_{n',l',m'} '
\frac{V_{0,0,0;n,l,m} V_{n,l,m;n',l',m'} V_{n',l',m';0,0,0} }{\left[
E_{0,0,0} - E_{n,l,m} \right] \left[
E_{0,0,0} - E_{n',l',m'} \right]}
+ \ldots~,
\end{eqnarray}
the denominators containing the energy eigenvalues of the harmonic oscillator
\begin{eqnarray}
\label{GE}
E_{n,l,m} = \hbar \omega \left( 2 n + l + \frac{3}{2} \right) \, .
\end{eqnarray}
The primed summations in (\ref{XO}) run over all values of the
quantum numbers $n,l = - \infty , \ldots , + \infty$ and
$m = - l , \ldots , + l$, excluding those for which the denominators vanish.
For the first three
orders we obtain from (\ref{MAA})--(\ref{GE})
\begin{eqnarray}
\label{PEOR}
E_{{\rm aux}} (\omega )
= \frac{3}{2} \hbar \omega - \frac{2}{\sqrt{\pi}}
\sqrt{\frac{M\omega}{\hbar}} e^2
- \frac{4}{\pi} \left( 1 +
\log  2 - \frac{\pi}{2} \right) \frac{M}{\hbar^2} e^4
- c \,\, \sqrt{\frac{M^3}{\hbar^7 \omega}} e^6 + \ldots
\end{eqnarray}
with the constant
\begin{eqnarray}
 c =  \frac{1}{\pi^{3/2}} & & \left\{ \sum_{n = 1}^{\infty}
\frac{1 \cdot 3 \cdot \ldots \cdot ( 2 n - 1 )}{2 \cdot 4 \cdot \ldots
\cdot 2 n} \, \frac{1}{n^2 \, ( n + 1/2 )}
- \sum_{n = 1}^{\infty}
\sum_{n' = 1}^{\infty}
\frac{1 \cdot 3 \cdot \ldots \cdot ( 2 n - 1 )}{2 \cdot 4 \cdot \ldots
\cdot 2 n}
\right. \nonumber \\
&& \times \left.
\frac{1 \cdot 3 \cdot \ldots \cdot ( 2 n' - 1 )}{2 \cdot 4 \cdot \ldots
\cdot 2 n'}
\frac{\phantom{}_3 F_2 \left( - n' , l + 1 , \frac{1}{2} ; l + \frac{3}{2} ,
\frac{1}{2} - n ; 1 \right)}{n \, n' \, ( n + 1/2 )}
\right\}
\approx 0.0318 \, .
\end{eqnarray}
The variational reexpansion procedure
described after Eq.~(\ref{V3}) replaces
$ \omega$ in the first term by $ \Omega(1-1)^{1/2}$, to be
expanded in the second $1$ up to third order as $1 - 1 / 2 -
1 / 8 - 1/16 = 5 / 16$. Correspondingly the term $3 \omega/2$ becomes
$15/32 \omega$.
The factor $ \omega^{1/2}$ goes over into $ \Omega^{1/2}(1-1)^{1/4}$,
to be  expanded
to second order in the second $1$,
yielding
$ \Omega^{1/4} ( 1 - 1/4 - 3 / 32 ) = 21 / 32$.
The next term in (\ref{PEOR}) happens to be independent
of $ \omega$ and needs no reexpansion, whereas the
last term remains unchanged since it is of highest order.
In this way we obtain from (\ref{PEOR}) the
third-order variational expression
\begin{eqnarray}
E^{(0)}_3
( \Omega ) = \frac{15}{32} \hbar \Omega - \frac{21}{16 \sqrt{\pi}}
\sqrt{\frac{M \Omega}{\hbar}} e^2 - \frac{4}{\pi} \left( 1 + \log  2 -
\frac{\pi}{2} \right) \frac{M}{\hbar^2} e^4 - c \,\,
\sqrt{\frac{M^3}{\hbar^7
\Omega}} e^6 \, . \label{E3}
\end{eqnarray}
%
%
%
We are now ready to optimize
successively the expansions
of first, second, and third order
 (\ref{E1}), (\ref{E2}) and
(\ref{E3}) with respect to the trial frequency
$\Omega$. From the extrema
we find the
frequencies
\begin{eqnarray}
\Omega_1 = \Omega_2 = \frac{16}{9 \pi} \frac{M e^4}{\hbar^3} \, ,
\hspace*{1cm} \Omega_3 = {c}'^{\,2} \, \frac{M e^4}{\hbar^3}
\end{eqnarray}
where ${c}' \approx 0.7254$
is the largest of the three
solutions
of the cubic equation
$15c'^{\,3}-{21}c'^{\,2}/{\pi}+16c.$
The corresponding
approximations to the ground state energy are
\begin{eqnarray}
E_N^{(0)} ( \Omega_N ) = - \gamma_N \frac{M e^4}{\hbar^2} \, ,
\end{eqnarray}
with the constants
\begin{eqnarray}
\label{VAL}
\gamma_1  =  \frac{4}{3 \pi} \approx 0.424 \, ,  \hspace*{1cm}
\gamma_2  =  \frac{5 + 4 \log  2}{\pi} - 2 \approx 0.474 \, , \hspace*{1cm}
\gamma_3  \approx  0.490,
\end{eqnarray}
which
quickly approaching the exact value $\gamma_{\rm ex} = 0.5$,
as shown in Fig.~3.
\begin{figure}[htb]
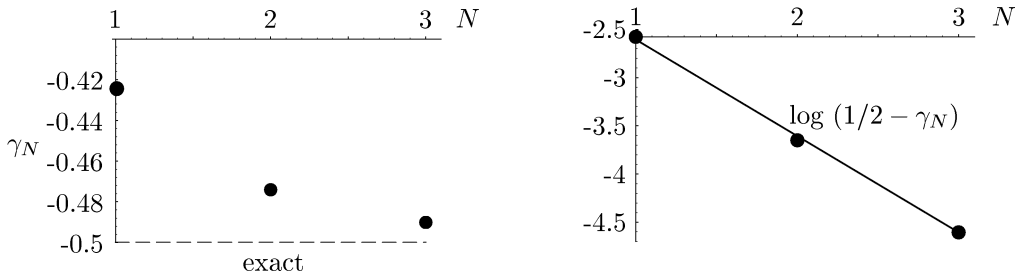

\input conv12.tps ~\\
\caption[]{Rapid approach of the variational
approximations to the ground state energy $E_N^{(0)} ( \Omega_N )$
to the correct ground state energy $-0.5$ (in natural energy units ${M e^4}/{\hbar^2} $).
The right-hand logarithmic plot shows a slope $-1$.}
\label{@K}
\end{figure}
\section{Summary and Outlook}

In this work we have
extended the rules for calculating higher orders
in
variational perturbation theory
from polynomial to nonpolynomial interactions.
The
effective classical potential of a quantum mechanical
system
is obtained from an extension of the known first-order
smearing formula,
and involves
certain convolutions with Gaussian functions. As an example,
we have applied the higher-order smearing formula to the
Coulomb system. We have illustrated the fast convergence of
the variational perturbation expansion
even  for such a singular potential.
The new
smearing formula
will help improving
the existing first-order
variational results
for partition function and density matrix,
also in dissipative
quantum systems \cite{Weiss,Vaia}. It will also be of use in treating
field theories with nonpolynomial interactions such as
Sine-Gordon and Liouville theories.

\section{Acknowledgement}

The authors are grateful to M. Bachmann and I. Mustapi\'c
for many stimulating discussions,
and to the Deutsche Forschungsgemeinschaft
for partial support under the grant Kl-256/25-1.
\end{document}